\documentclass[aps,pra,twocolumn,amsmath,amssymb, superscriptaddress,tightenlines]{revtex4}
\usepackage{graphicx}% Include figure files
\usepackage{epsfig}
\usepackage{dcolumn}% Align table columns on decimal point
\usepackage{bm}% bold math
\usepackage{amssymb}
\usepackage{bm}
\usepackage{amsmath, bm}
\usepackage{upgreek}
\usepackage{subfigure}
\usepackage[colorlinks,citecolor=blue,linkcolor=blue,hyperindex]{hyperref}

\begin{document}

\title{Multiphoton resonance and chiral transport in the generalized Rabi model}
\author{Ken K. W. Ma}
\affiliation{Department of Physics, Brown University, Providence, Rhode Island 02912, USA}
\affiliation{National High Magnetic Field Laboratory, Tallahassee, Florida 32310, USA}
\date{\today}

%----------------------- Abstract ------------------------------

\begin{abstract}
The generalized Rabi model (gRM) with both one- and two-photon coupling terms has been successfully implemented in circuit quantum electrodynamics systems. In this paper, we examine theoretically multiphoton resonances in the gRM and derive their effective Hamiltonians. We show that all three- to six-photon resonances can be achieved by involving two intermediate states with different detunings in the system. Furthermore, we study the interplay between multiphoton resonance and chiral transport of photon Fock states in a resonator junction with broken time-reversal symmetry. Depending on the qubit-photon interaction and photon-hopping amplitude, we find that the ultrastrong coupling system can demonstrate different short-time dynamics.
\end{abstract}

\maketitle

%-------------------------- Introduction -----------------------------

\section{Introduction} 

A two-level atom interacting with a quantized electromagnetic field is one of the most basic and oldest problems in quantum optics. The system is described by the celebrated quantum Rabi model~\cite{Rabi1937}. When the atom-photon interaction is weak, the rotating-wave approximation (RWA) applies and leads to the Jaynes-Cummings model (JCM)~\cite{Jaynes-Cummings}. Although counter-rotating (excitation-number-nonconserving) processes are eliminated by the RWA, the JCM had been a widely accepted model in quantum optics for a long time~\cite{Scully_book}. Recent advances in fabricating circuit quantum electrodynamics (cQED) systems~\cite{cQED-review} make it feasible for the atom-photon interaction to reach the ultrastrong coupling (USC)~\cite{Nature-USC-review, PRL-USC} or even the deep-strong-coupling (DSC) regimes~\cite{PRL-DSC, PRX-DSC, Yoshihara-DSC-exp}. Since the RWA breaks down in these regimes, it has reignited a large amount of studies of the original Rabi model. The model was shown to be quantum integrable~\cite{Braak-solution, Batchelor-Zhou2015}, and attracted various attempts to construct its analytic solutions~\cite{Jia-solution, Braak-solution, Chen-PRA2012, Moroz2013, Lee-solution, Macieijewski-JPA, Lee2017}. On a more physical perspective, previous works have shown that counter-rotating terms could lead to many novel phenomena, and may find different applications that are absent in the JCM~\cite{Nori_nature_review, RMP-review, Saidi-EPJD, Ma-Law2015, Nori-multiphoton, Garziano-PRL2016, Nori-nonlinear, Nori-no-photon, Nori-frequency, Garziano-bunching, Zhao-qutrits, Wu_opt-express, Wu-PRA2018, Nori-dissipation, Nori-Bell-GHZ, Qi-Jing_NOON, Wu-PRA2019, Dodonov-PRA, Nori-optomechanics, Chen-probe, Ridolfo-blockade, Ridolfo-radiation, Nori-nonclassical, Ashhab-superradiance, DeLiberato-vacuum, Nation-vacuum, Stassi-vacuum, Garziano-vacuum2013, Garziano-vaccum2014, Law-PRA2014, Ai-anti-Zeno, Cao-Zeno, Kyaw-scalable}. 

Aside from the original Rabi model, a two-photon coupling term with a two-level system was successfully simulated in both cQED~\cite{Felicetti-collapse, Puebla-2ph-Rabi} and cold-atom setups~\cite{cold-atom-Rabi}. This nonlinear term may lead to squeezing of light~\cite{Peng-2ph-Rabi, CFLo-2ph-Rabi} and interaction-induced spectral collapse~\cite{Felicetti-collapse, Chen-2ph-collapse, Lupo-2ph-collapse, Rico-collapse2019, Cong-polaron-collapse}. Furthermore, a generalized Rabi model (gRM) with both one-photon and two-photon terms has emerged in the last decade~\cite{exp-GRM1, exp-GRM2, footnote-Rabi-Stark}. It was found that both terms are tunable and the two-photon term need not be suppressed by enhancing the single-photon term~\cite{SPP-TPP-tune}. Mathematical consequences of the broken $\mathbb{Z}_2$ symmetry and analytic solutions of the model have been well documented in recent works~\cite{gRM-solution-PRA, gRM-solution-Sci}. Also, it is exciting that the gRM may open a door to simulate particle dynamics in (1+1)-dimensional curved spacetime~\cite{Pedernales-PRL2018}. Meanwhile, not much is known for other physical effects of the generalized model. We believe it is tempting to explore new features in the gRM and examine their corresponding physical applications.

In this paper, our first goal is examining theoretically multiphoton resonances in the gRM. With different detunings between atomic transition frequency and photon frequency, we show that all three- to six-photon resonances can be achieved. Also, we derive effective Hamiltonians describing these resonances. The result is valuable because resonances involving even numbers of photons are forbidden in the original Rabi model due to its $\mathbb{Z}_2$ symmetry~\cite{Benivegna-Z2, Braak-solution}. It is remarked that a parity-violating coupling between the resonator mode and the flux qubit exists in a realistic cQED setup~\cite{Nature-USC-review} and its related asymmetric Rabi model~\cite{Braak-solution, Chen-PRA2012, Larson2013}. Nevertheless, the absence of a two-photon term there implies that an $N$-photon resonance must involve at least $N-1$ intermediate states. This kind of process is suppressed even in the USC regime. In other words, the gRM provides an opportunity to realize multiphoton resonances involving large numbers of photons in a more effective way. This is further elaborated in Sec.~\ref{sec:NOON}, and has important consequences in NOON-states generation via adiabatic passage~\cite{Ma-Law2015}.

The above results lead us to the second goal of the paper: studying the interplay between multiphoton resonances and chiral photon transfer in cQED lattices. Although breaking time-reversal symmetry to achieve chiral transport of photons is challenging~\cite{Solijacic-QHE-light, Haldane-QHE-light, Girvin-TRS, Girvin-JC, Carusotto2011, Hafezi2014, Hey-Li2018, focus-NJP, nature-topo-optics, RMP-topo-optics}, an accessible scheme was proposed for cQED systems~\cite{Girvin-TRS}. Specifically, synthetic gauge flux for photons can be induced by inserting Josephson rings to the system as on-chip circulators. When the flux is tuned properly, photon transfer with a preferred chirality was realized in a junction of three microwave resonators~\cite{Girvin-TRS}. A similar effect can also occur when each resonator couples to a superconducting qubit~\cite{Girvin-JC}. The result has opened up the possibility of simulating novel phases of photons in cQED systems. In this paper, we revisit the coupled system without employing the RWA and without limiting to one-photon transfer. Each coupled qubit-resonator in our system may achieve multiphoton resonance. However, this effect needs to compete with photon hopping between different resonators. When the photon-hopping strength and the effective multiphoton-resonant coupling strength are tuned, it is expected that the system will exhibit different types of short-time dynamics.

The paper is organized as follows. In the first part of our work, we begin by reviewing the generalized Rabi model and outlining the perturbation theory for analyzing multiphoton resonances in Sec.~\ref{sec:Hamiltonian-GRM}. Then, we examine three-photon resonance in Sec.~\ref{sec:3-photon}, and compare the results with the original Rabi model. In Sec.~\ref{sec:4-6-ph-resonance}, we demonstrate the possibility of achieving four- to six-photon resonances in the gRM. In the second part of the paper, we study in Sec.~\ref{sec:topo} the interplay between chiral photon transport and multiphoton resonance in cQED systems. By reducing the ratio of photon hopping strength to multiphoton-resonant coupling strength, the transition from a chiral photon transfer to a suppression of photon hopping in the short-time dynamics of the system is illustrated. Last, we conclude our work in Sec.~\ref{sec:conclusion}.

\section{Review of the generalized Rabi model and perturbation theory} 
\label{sec:Hamiltonian-GRM}

The generalized Rabi model with both one-photon and two-photon coupling terms is described by the Hamiltonian:
\begin{align} \label{eq:Hamiltonian-gRM}
\nonumber
H
=&~\frac{\omega_a}{2} \sigma_z
+\omega_c a^\dagger a
+\lambda \left(a+a^\dagger\right)\sigma_x
\\
&+\kappa\left[a^2+\left(a^\dagger\right)^2\right]\sigma_x.
\end{align}
In this paper, we set $\hbar=1$ unless specified. The transition frequency of the two-level atom is denoted by $\omega_a$. A photon with frequency $\omega_c$ is annihilated (created) by the operator $a$ $(a^\dagger)$. We denote the ground state and the excited state of the two-level atom as $\left|g\rangle\right.$ and $\left|e\rangle\right.$, respectively. Then, the Pauli matrices are given by $\sigma_z=|e\rangle\langle e|-|g\rangle\langle g|$ and 
$\sigma_x=|e\rangle\langle g|+|g\rangle\langle e|$. The one- and two-photon coupling terms with the atom have coupling strengths $\lambda$ and $\kappa$, respectively. 

In the absence of the two-photon coupling term (i.e., when $\kappa=0$), it is possible to define a parity operator for the quantum Rabi model~\cite{Benivegna-Z2, Braak-solution}:
\begin{eqnarray}
\Pi_1=\exp{\left[i\pi \left(\frac{1+\sigma_z}{2} + a^\dagger a\right)\right]}.
\end{eqnarray}
The parity operator satisfies $\left[H, \Pi_1\right]=0$. By acting $\Pi_1$ on the bare states 
$\left|g,n\rangle\right.$ and $\left|e,n\rangle\right.$, it can only take eigenvalues of $\pm 1$. Similarly, one may define a corresponding $\mathbb{Z}_4$-symmetry operator for the two-photon Rabi model (i.e., when $\lambda=0$)~\cite{Emary-Z4-operator}:
\begin{eqnarray}
\Pi_2=\exp{\left[i\pi \left(\frac{1+\sigma_z}{2} + \frac{a^\dagger a}{2}\right)\right]}.
\end{eqnarray}
Every eigenstate of the Hamiltonian must be a linear superposition of bare states in the same subspace with the same eigenvalue of $\Pi_1$ or $\Pi_2$. However, the discrete symmetry is broken in the generalized Rabi model with both one- and two-photon terms. Thus it becomes more difficult to study the analytic solutions of the model. At the same time, it opens the door to other multiphoton resonances, which are forbidden in the original Rabi model. We review the third-order perturbation theory for multiphoton resonances in a largely detuned Rabi model.

\subsection{Perturbation theory for multiphoton resonance}
\label{sec:eqn-resonance}

In this paper, we study multiphoton resonances between two bare states $|i\rangle$ and 
$|f\rangle$. We only focus on resonances which involve two intermediate states. The two bare states are $|i\rangle=|g,n_0+n\rangle$ and $|f\rangle=|e, n_0\rangle$. In the absence of atom-photon interaction, these two states are degenerate when $\omega_c=\omega_a/n$. An effective Hamiltonian to describe the resonance can be obtained by eliminating the intermediate states. This can be done by several approaches, such as adiabatic elimination~\cite{Ma-Law2015}, the generalized James' effective Hamiltonian approach~\cite{effective-James}, and third-order perturbation theory~\cite{Nori-nonlinear}. The derivation based on the second approach is discussed in Appendix~\ref{app:effective-H}. The end result is in the following form:
\begin{align} \label{eq:reduced-H}
\nonumber
H^{\text{$n$-ph}}_{\text{eff}}
=&~\left(E_i+\Delta E_i\right)|i\rangle \langle i|
+\left(E_f+\Delta E_f\right)|f\rangle \langle f|
\\
&+\Omega_{\text{eff}}^{\text{$n$-ph}}
\left(|i\rangle \langle f| + |f\rangle \langle i|  \right).
\end{align}
Here, the unperturbed energy for the state $|i\rangle$ is denoted by $E_i$. Due to the atom-photon interaction, the energy levels are Stark shifted. Consequently, the required photon frequency of achieving the $n$-photon resonance (or equivalently, the resonant frequency) is perturbed away from $\omega_c=\omega_a/n$. The perturbed resonant frequency $\omega_c'$ can be obtained by equating the diagonal elements of $H^{\text{$n$-ph}}_{\text{eff}}$, that is solving $E_i+\Delta E_i=E_f+\Delta E_f$. At the resonance, the two nearly degenerate energy levels develop an avoided crossing with a gap $2|\Omega_{\text{eff}}^{\text{$n$-ph}}|$.

In the following discussion, we directly use the result from second-order perturbation theory to obtain the leading order terms in $\Delta E_i$:
\begin{eqnarray} \label{eq:Stark-shift}
\Delta E_i
=
\sum_{\alpha}
\frac{\left|\langle \alpha\left| V\right| i\rangle\right|^2}{E_i-E_\alpha}.
\end{eqnarray}
Here, the symbol $V$ denotes the atom-photon interaction terms in the gRM, i.e., 
$V=\lambda(a+a^\dagger)\sigma_x + \kappa(a^2+\left.a^\dagger\right.^2)\sigma_x$. All intermediate bare states which can be connected to $|i\rangle$ by $V$ are labeled as
$|\alpha\rangle$. The result of $\Delta E_i/\omega_a$ and $\omega_c'/\omega_a$ from Eq.~\eqref{eq:Stark-shift} will be accurate to the order of $(\lambda/\omega_a)^2$ and 
$(\kappa/\omega_a)^2$. For the effective coupling strength $\Omega_{\text{eff}}^{\text{$n$-ph}}$, it can be deduced from the third order perturbation theory~\cite{Nori-nonlinear}:
\begin{eqnarray} \label{eq:perturb-eff}
\Omega_{\text{eff}}^{\text{$n$-ph}}
=\sum_{\alpha, \beta}
\frac{\langle f\left| V\right|\beta\rangle
\langle \beta\left| V\right|\alpha\rangle
\langle \alpha\left| V\right| i\rangle}
{\left(E_i-E_\alpha\right)\left(E_i-E_\beta\right)}.
\end{eqnarray}
The symbol $|\beta\rangle$ denotes another intermediate state. All results obtained from Eqs.~\eqref{eq:Stark-shift} and~\eqref{eq:perturb-eff} are consistent with the generalized James' effective Hamiltonian approach discussed in Appendix~\ref{app:effective-H}.

\section{Three-photon resonance in the generalized Rabi model}
\label{sec:3-photon}

\begin{figure}
\includegraphics[width=3.0in]{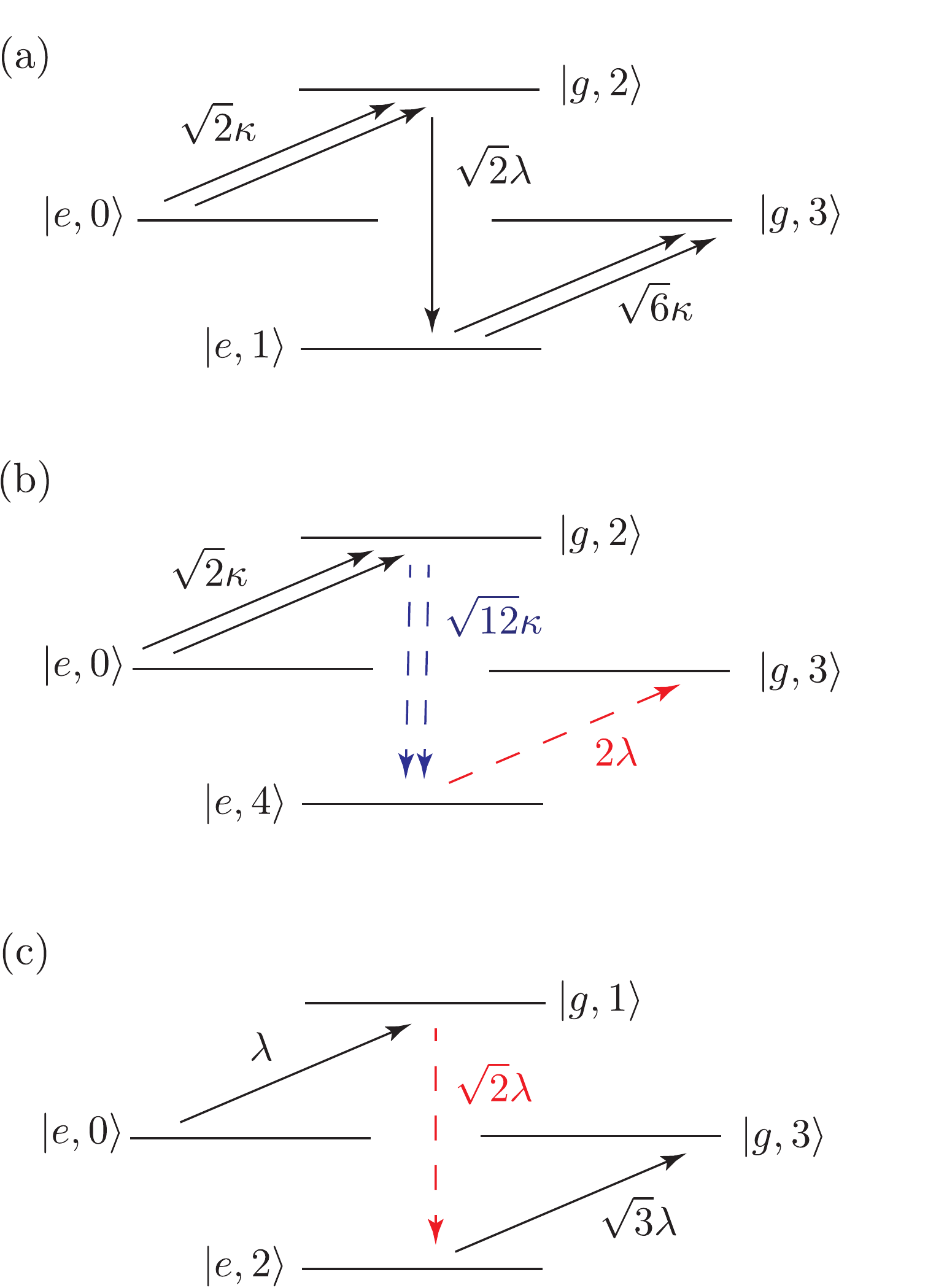}
\caption{(Color online) Three different coupling schemes for the three-photon resonance in the gRM in Eq.~\eqref{eq:Hamiltonian-gRM}. Here, counter-rotating processes are labeled by dashed lines. Single (double) lines with arrows indicate coupling between the bare states by the single-photon (two-photon) term in the generalized Rabi model. In contrast to the original Rabi model (i.e., when $\kappa=0$), three-photon resonance is possible even when the RWA applies as illustrated in (a). The coupling scheme shown in (c) was introduced in Ref.~\cite{Ma-Law2015}. The values near the arrowed lines are the matrix elements $\langle\alpha_1, n_1\left| V\right| \alpha_2, n_2\rangle$.}
\label{fig:three-photon-resonance}
\end{figure}

In the original Rabi model (i.e., $\kappa=0$), a three-photon resonance can occur when the frequency of the photon field is tuned to $\omega_c\approx\omega_a/3$~\cite{Ma-Law2015}. Physically, this is possible for two reasons. First, the two bare states 
$\left|g,3\rangle\right.$ and $\left|e,0\rangle\right.$ are nearly degenerate. Second, these two bare states can be connected by intermediate states via counter-rotating processes. Thus an avoided crossing is formed between the two energy levels. Away from the avoided crossing, the corresponding eigenvectors of these two levels have high overlaps with the two bare states. In the framework of the generalized Rabi model, we revisit the three-photon resonance. The three different schemes for achieving the resonance in the gRM are illustrated in Fig.~\ref{fig:three-photon-resonance}.

\subsection{General consideration without rotating-wave approximation} 

For simplicity, we only consider the three-photon resonance between the bare states $|g,3\rangle$ and $|e,0\rangle$ in detail. It is straightforward to generalize the discussion to any pair of $|g,n_0+3\rangle$ and $|e,n_0\rangle$. To the leading order, the one-photon and two-photon terms in the gRM couple $|g,3\rangle$ to $|e,5\rangle$, $|e,4\rangle$, $|e,2\rangle$, and $|e,1\rangle$. Similarly, $|e,0\rangle$ is coupled to $|g,2\rangle$ and $|g,1\rangle$. From the discussion in Sec.~\ref{sec:eqn-resonance}, the resonant frequency is determined as
\begin{eqnarray} \label{eq:freq-3-ph}
\frac{\omega_c'}{\omega_a}
=\frac{1}{3}
+3\left(\frac{\lambda}{\omega_a}\right)^2
+12\left(\frac{\kappa}{\omega_a}\right)^2.
\end{eqnarray}
The same result can be obtained from Eq.~\eqref{eq:general-freq} with $n=3$ and $n_0=0$. By employing Eq.~\eqref{eq:perturb-eff} and summing the contributions from all three possible coupling schemes in Fig.~\ref{fig:three-photon-resonance}, we obtain the effective coupling strength for the three-photon resonance:
\begin{eqnarray} \label{eq:split-3-ph}
\Omega_{\text{eff}}^{\text{3-ph}}
=-\left(\frac{27\sqrt{6}\lambda\kappa^2}{\omega_a^2}
+\frac{9\sqrt{6}\lambda^3}{4\omega_a^2}\right).
\end{eqnarray}
Note that the second term in the final result agrees with the result for three-photon resonance in the original Rabi model~\cite{Ma-Law2015}. We compare the results from perturbation theory and numerical diagonalization. 

\begin{figure}
\includegraphics[width=3.3in]{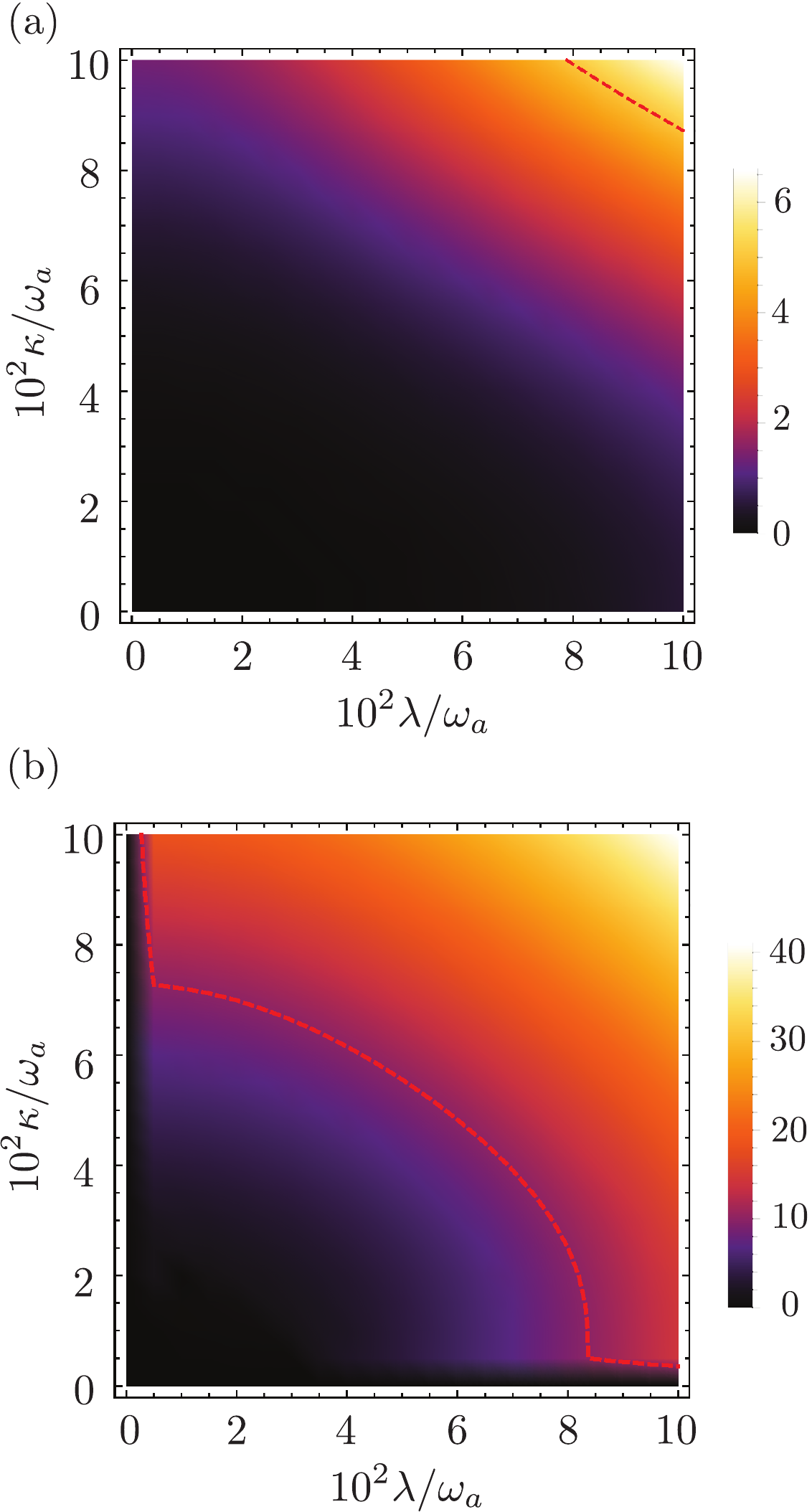}
\caption{(Color online) Percentage errors between (a) resonant frequencies $\omega_c'/\omega_a$ and (b) energy splittings $\Delta/\omega_a$ for the three-photon resonance from perturbation theory and numerical diagonalization. For reference, we mark the regions (below the red dashed lines) in (a) and (b) that have percentage errors less than 
$5\%$ and $10\%$, respectively.}
\label{fig:3-ph-comparison}
\end{figure}

\begin{figure}
\includegraphics[width=3.3in]{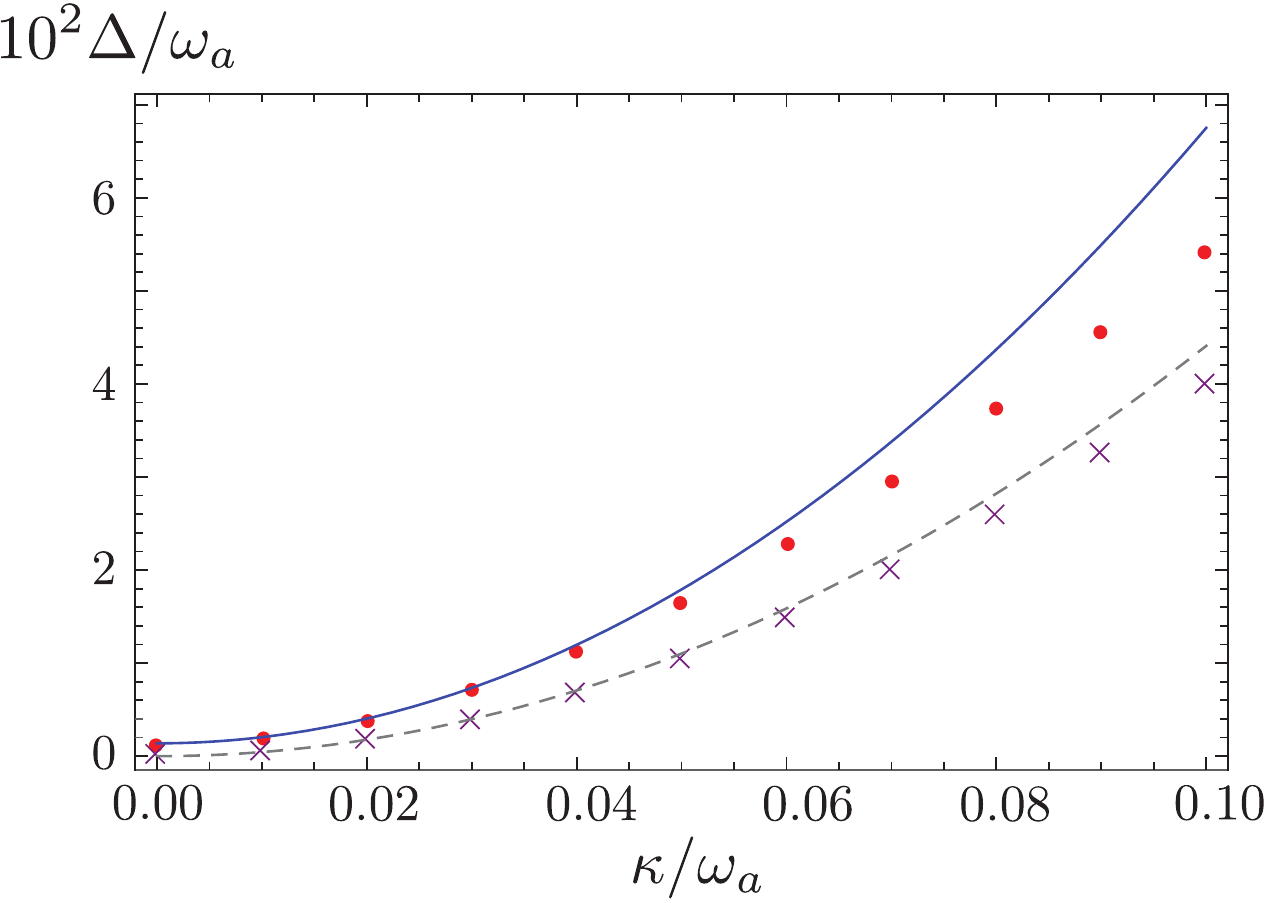} 
\caption{(Color online) The energy splitting at the three-photon resonance as a function of $\kappa/\omega_a$. Here, we fix $\lambda/\omega_a=0.05$. The results for the general case without the RWA are denoted by a blue solid line (perturbation theory) and a red dotted line (numerical diagonalization). The other two lines present results with the application of the RWA. The gray dashed line and purple crosses denote results from perturbation theory and numerical simulation, respectively. }
\label{fig:3-ph-split}
\end{figure}

In Fig.~\ref{fig:3-ph-comparison}, we show the percentage difference between the resonant frequency from Eq.~\eqref{eq:freq-3-ph} and numerical diagonalization. Also, we obtain numerically the three-photon Rabi splitting at the resonance, i.e., $\Delta=2\left|\Omega_{\text{eff}}^{3-\text{ph}}\right|$, and its percentage difference from Eq.~\eqref{eq:split-3-ph}. For $\lambda/\omega_a<0.1$ and $\kappa/\omega_a<0.1$, the approximate result for the resonant frequency has a percentage error of $5\%$ or smaller in most of the region. At the same time, there is a considerable region where the approximate energy splitting has a percentage error smaller than $10\%$. For another illustration, we plot $\Delta/\omega_a$ as a function of $\kappa/\omega_a$ with $\lambda/\omega_a=0.05$ in Fig.~\ref{fig:3-ph-split}. For comparison, both results from numerical simulation and Eq.~\eqref{eq:perturb-eff} are included. 

\subsection{Rotating-wave approximation} 

When the atom-photon interaction is weak, the RWA leads to the Jaynes-Cummings-type Hamiltonian:
\begin{align}
\nonumber
H_{\text{RWA}}
=&~\frac{\omega_a}{2} \sigma_z
+\omega_c a^\dagger a
+\lambda \left(a\sigma^{+} + a^\dagger \sigma^{-} \right)
\\
&~+\kappa\left(a^2\sigma^{+} +\left. a^\dagger\right.^2 \sigma^{-} \right).
\end{align}
Here, we define the symbols $\sigma^{+}=\left|e\rangle\langle g\right|$ and
$\sigma^{-}=\left|g\rangle\langle e\right|$. From Fig.~\ref{fig:three-photon-resonance}(a), it is observed that a three-photon resonance can also occur in the gRM when the RWA applies. This is drastically different from the situation in the original Rabi model, where the resonance can only happen with the presence of counter-rotating terms~\cite{Ma-Law2015}. From Eq.~\eqref{eq:perturb-eff}, we deduce the effective coupling strength of the three-photon resonance under the RWA as
\begin{eqnarray}
\Omega_{\text{RWA, eff}}^{3-\text{ph}}
=-\frac{18\sqrt{6}\kappa^2\lambda}{\omega_a^2}.
\end{eqnarray}
The corresponding resonant frequency is
\begin{eqnarray} \label{eq:freq-3-RWA}
\left(\frac{\omega_c'}{\omega_a}\right)^{\text{3-ph}}_{\text{RWA}}
=\frac{1}{3}+2\left(\frac{\lambda}{\omega_a}\right)^2
+8\left(\frac{\kappa}{\omega_a}\right)^2.
\end{eqnarray}

\section{Multiphoton resonances with four to six photons} 
\label{sec:4-6-ph-resonance}

The additional two-photon term in the generalized Rabi model leads to the possibility of realizing multiphoton resonances with larger numbers of photons. By limiting to resonances involving two intermediate states, four- to six-photon resonances can be achieved.

\subsection{Four-photon resonance} 

\begin{figure}
\includegraphics[width=3.0in]{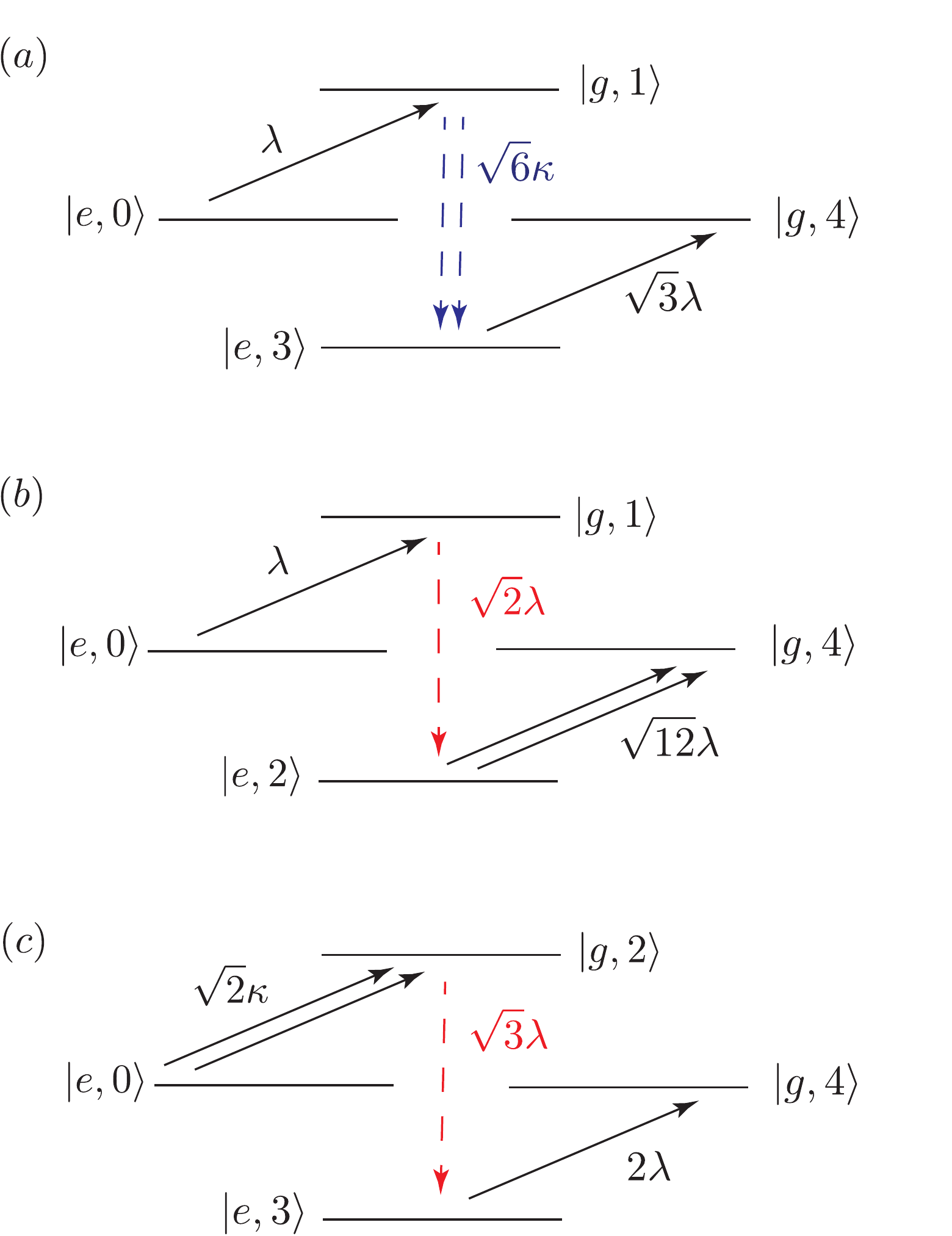}
\caption{(Color online) Three different coupling schemes for the four-photon resonance in the gRM in Eq.~\eqref{eq:Hamiltonian-gRM}. Here, counter-rotating processes are labeled by dashed lines. Single (double) lines with arrows indicate coupling between the bare states by the single-photon (two-photon) term in the generalized Rabi model. The values near the arrowed lines are the matrix elements $\langle\alpha_1, n_1\left| V\right| \alpha_2, n_2\rangle$.}
\label{fig:four-photon-resonance}
\end{figure}

\begin{figure} [htb]
\includegraphics[width=3.3in]{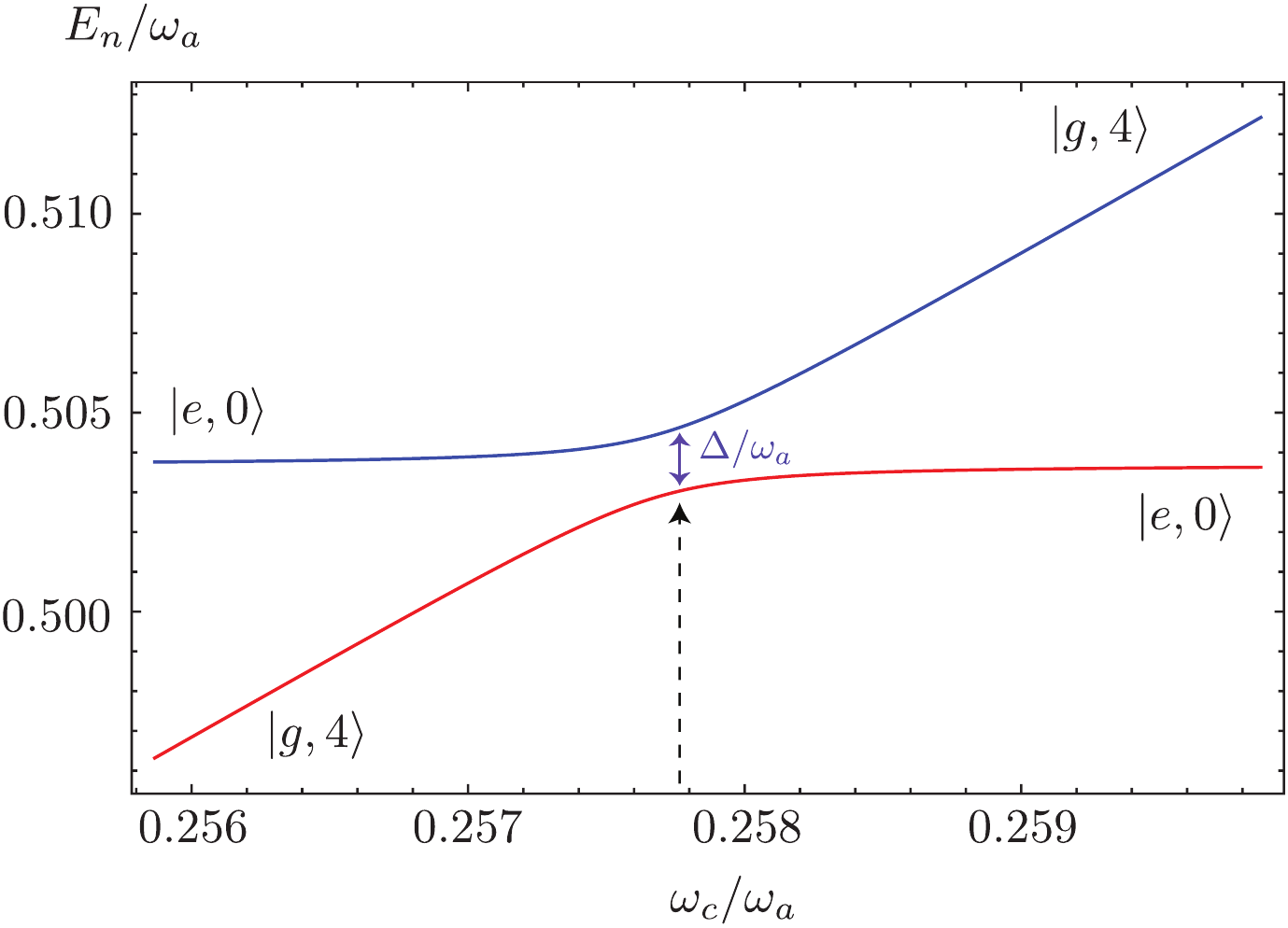}
\caption{(Color online) An avoided crossing in the energy levels of the fourth and fifth excited states in the gRM when $\omega_c/\omega_a\approx 1/4$. The Dirac kets show the major components of the eigenstates away from the four-photon resonance. The resonant frequency is determined numerically as $\omega_c'/\omega_a\approx 0.258$, with a corresponding energy splitting $\Delta/\omega_a\approx 1.57\times 10^{-3}$. Here, we set $\lambda/\omega_a=0.05$ and $\kappa/\omega_a=0.01$. Note that both $\lambda$ and $\kappa$ must be nonzero to achieve the resonance.}
\label{fig:4-ph-av-crossing}
\end{figure}

\begin{figure} [htb]
\includegraphics[width=3.3in]{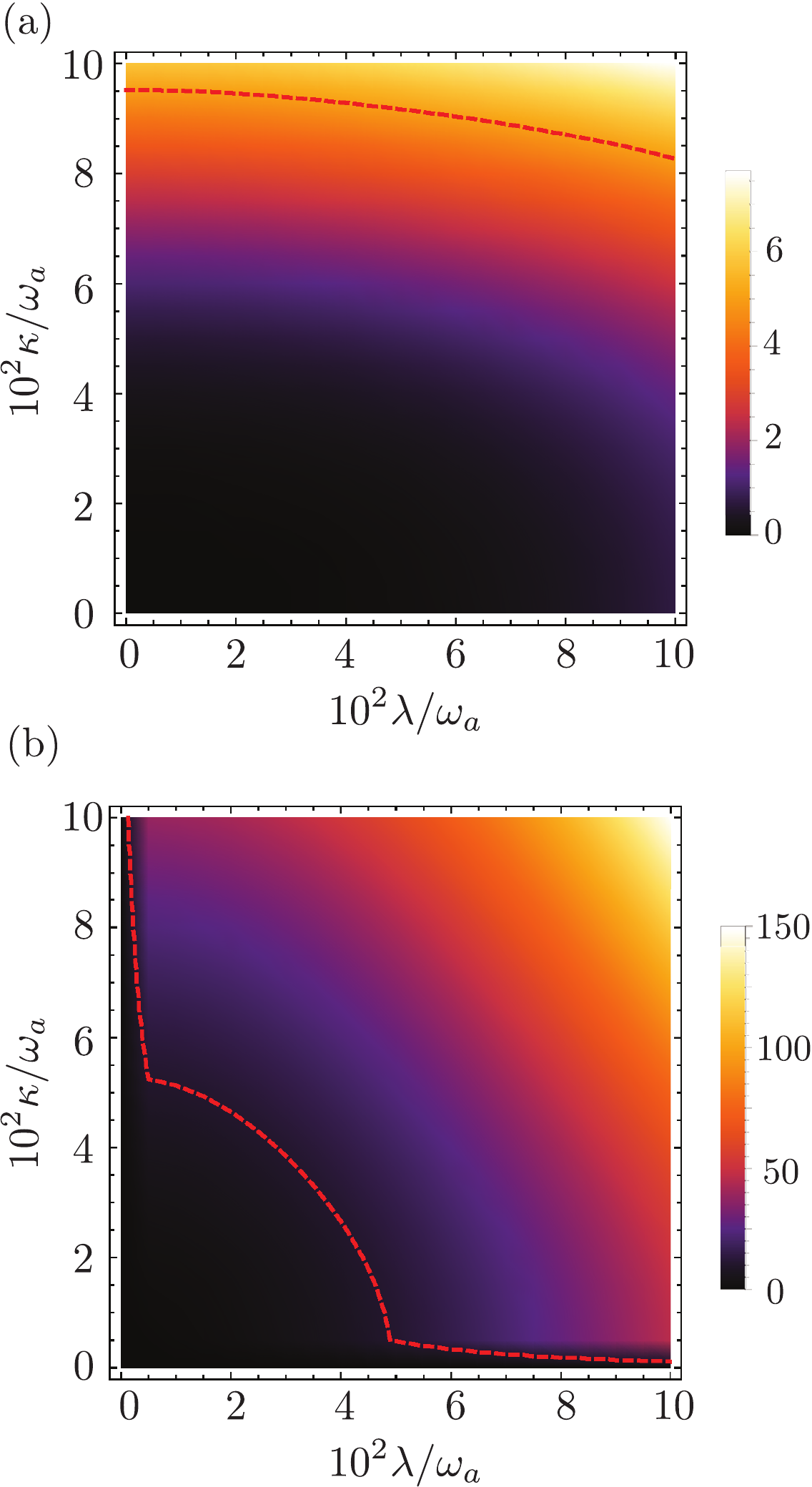}
\caption{(Color online) Percentage errors between (a) the resonant frequencies $\omega_c'/\omega_a$ and (b) energy splitting $\Delta/\omega_a$ for the four-photon resonance from perturbation theory and numerical diagonalization. For reference, we mark the regions (below the red dashed lines) in (a) and (b) that have percentage errors less than $5\%$ and $10\%$, respectively.}
\label{fig:4-ph-comparison}
\end{figure}

We begin by considering the four-photon resonance between the bare states $|g,4\rangle$ and $|e,0\rangle$. The resonance is interesting for two reasons. First, the two bare states belong to different symmetry classes under $\Pi_1$ and $\Pi_2$. Thus it is impossible to realize the resonance in the Rabi model with only one- or two-photon terms. A possible solution is introducing a parity-violating term in the Hamiltonian. This is achievable in circuit QED systems~\cite{Nature-USC-review}. Nevertheless, a higher-order coupling involving three intermediate states is required. The second feature of the four-photon resonance is the necessity of having counter-rotating terms if \textit{only two intermediate states} are involved. This is demonstrated in the three different coupling schemes in Fig.~\ref{fig:four-photon-resonance}. Using Eqs.~\eqref{eq:Stark-shift} and~\eqref{eq:perturb-eff}, we determine the resonant frequency 
\begin{eqnarray}
\left(\frac{\omega_c'}{\omega_a}\right)^{\text{$4$-ph}}
=\frac{1}{4}
+\frac{8}{3}\left(\frac{\lambda}{\omega_a}\right)^2
+12\left(\frac{\kappa}{\omega_a}\right)^2.
\end{eqnarray}
and the effective coupling as
\begin{eqnarray}
\Omega_{\text{eff}}^{\text{4-ph}}
=-\frac{128\sqrt{6}}{9}\left(\frac{\lambda}{\omega_a}\right)^2 \kappa.
\end{eqnarray}
In Fig.~\ref{fig:4-ph-av-crossing}, we plot the energies of fourth excited and fifth excited states of the gRM at $\omega_c\approx\omega_a/4$, $\lambda/\omega_a=0.05$, and $\kappa/\omega_a=0.01$. Percentage differences between the results from perturbation theory and numerical diagonalization are shown in Fig~\ref{fig:4-ph-comparison}.

It is necessary to clarify that a four-photon resonance can also occur under the RWA, for example, via the leading-order coupling scheme: $|e,0\rangle\rightarrow|g,2\rangle\rightarrow |e,1\rangle\rightarrow|g,3\rangle\rightarrow|e,2\rangle\rightarrow|g,4\rangle$. However, this is a fifth-order process which is suppressed and cannot be effectively realized when $\lambda/\omega_a<0.1$ and $\kappa/\omega_a<0.1$.

\subsection{Five- and six-photon resonance} 

By involving two intermediate states, multiphoton resonances involving five and six photons can also be realized in the gRM. First, consider the five-photon resonance between 
$|e,0\rangle$ and $|g,5\rangle$. To the leading order, these two states can be connected in three different ways: 
\begin{eqnarray}
\begin{aligned}
&|e,0\rangle \rightarrow |g,1\rangle \rightarrow |e,3\rangle \rightarrow |g,5\rangle,
\\
&|e,0\rangle \rightarrow |g,2\rangle \rightarrow |e,3\rangle \rightarrow |g,5\rangle,
\\
&|e,0\rangle \rightarrow |g,2\rangle \rightarrow |e,4\rangle \rightarrow |g,5\rangle.
\end{aligned}
\end{eqnarray}
Using Eqs.~\eqref{eq:Stark-shift} and~\eqref{eq:perturb-eff}, it is straightforward to determine the resonant frequency 
\begin{eqnarray}
\left(\frac{\omega_c'}{\omega_a}\right)^{\text{$5$-ph}}
=\frac{1}{5}
+\frac{5}{2}\left(\frac{\lambda}{\omega_a}\right)^2
+\frac{40}{3}\left(\frac{\kappa}{\omega_a}\right)^2.
\end{eqnarray}
and the effective coupling strength
\begin{eqnarray}
\Omega_{\text{eff}}^{\text{5-ph}}
=-\frac{125\sqrt{30}\kappa^2\lambda}{9\omega_a^2}.
\end{eqnarray}

Similarly, the following leading-order coupling
\begin{eqnarray}
|e,0\rangle \rightarrow |g,2\rangle \rightarrow |e,4\rangle \rightarrow |g,6\rangle
\end{eqnarray}
leads to a possible six-photon resonance between $|e,0\rangle$ and $|g,6\rangle$. The corresponding resonant frequency and the effective coupling are determined as
\begin{eqnarray}
\left(\frac{\omega_c'}{\omega_a}\right)^{\text{$6$-ph}}
=\frac{1}{6}
+\frac{12}{5}\left(\frac{\lambda}{\omega_a}\right)^2
+15\left(\frac{\kappa}{\omega_a}\right)^2
\end{eqnarray}
and
\begin{eqnarray}
\Omega_{\text{eff}}^{\text{6-ph}}
=-\frac{27\sqrt{5}\kappa^3}{\omega_a^2}.
\end{eqnarray}
In principle, one can obtain similar plots for the percentage difference between the results from perturbation theory and numerical diagonalization. As the number of photons being involved increases, third-order perturbation theory is accurate in a much smaller region of the parameter space. Nevertheless, the multiphoton resonances do exist. At the same time, the perturbation theory approach is still valid given that the atom-photon interaction is sufficiently weak. 

\subsection{Possible application in NOON state generation}
\label{sec:NOON}

In previous works, different applications of multiphoton resonances have been proposed. Some examples include production of coherent photons~\cite{Ma-Law2015}, simultaneous excitation of several atoms~\cite{Garziano-PRL2016}, frequency conversion~\cite{Nori-frequency}, and preparation of different entangled photon states~\cite{Nori-Bell-GHZ, Qi-Jing_NOON}. Here, we want to address a possible application of our results in NOON state generation. The discussion follows our previous work closely~\cite{Ma-Law2015}. Such a discussion also allows us to highlight the advantage of realizing and exploiting multiphoton resonances in the generalized Rabi model.

Suppose we have two spatially separated qubits which have identical transition frequency, i.e., $\omega_a=\omega_{a,1}=\omega_{a,2}$. Each of them is coupled to its own microwave resonator. There is no coupling between different resonators or different qubits. We further assume the whole setup can be modeled by a sum of two copies of Eq.~\eqref{eq:Hamiltonian-gRM}, but with $\omega_c$ being time-dependent. Initially, the two qubits are prepared in the entangled state: $\left(|g,e\rangle+|e,g\rangle\right)/\sqrt{2}$. Meanwhile, both resonators have the same mode frequency and contain zero excitation. At the beginning, the mode frequency $\omega_c=\omega_{c,1}=\omega_{c,2}$ is smaller than the multiphoton resonant frequency $\omega_c'$. Following Ref.~\cite{Ma-Law2015}, we increase $\omega_c(t)$ adiabatically and finally reach $\omega_c(t_f)>\omega_c'$. After reaching 
$\omega_c(t_f)$, $\omega_c$ remains unchanged. In the ideal situation, the bare states 
$|e,0\rangle$ and $|g,0\rangle$ will evolve to $|g,N\rangle$ and $|g,0\rangle$, respectively. As a result, the two resonators will acquire a final state $|\text{NOON}\rangle=\left(|N, 0\rangle+e^{i\theta}|0, N\rangle\right)/\sqrt{2}$, whereas both qubits are in the ground state.

In fact, the above discussion assumed that the two nearly degenerate eigenstates of the gRM take the form $|\psi_{\pm}\rangle=(|e,0\rangle \pm |g,N\rangle)/\sqrt{2}$ at the $N$-photon resonance. This is a good approximation only when both $\lambda/\omega_a$ and $\kappa/\omega_a$ are sufficiently small. If this condition is violated, then $|\psi_{\pm}\rangle$ also contain significant projections along other bare states in addition to $|e,0\rangle$ and $|g,N\rangle$. In this scenario, the final state obtained from the adiabatic passage does not have a high overlap with $|\text{NOON}\rangle$. This issue makes it difficult to put our proposal in practice. For a linearly increasing $\omega_c(t)$, the transition across the multiphoton resonance can be approximated by a Landau-Zener transition~\cite{Ma-Law2015}. Within this approximation, the sweeping speed of $\omega_c(t)$ should be small compared to the energy splitting squared at the avoided crossing. At the same time, spontaneous decay of qubits (or two-level atoms) and photon leakage from resonators (or microwave cavities) occur in a real setup. Therefore, the sweeping speed must be small but simultaneously large enough to prevent significant damping effects. Our previous discussion shows that the gap $\Delta/\omega_a$ scales with the third order of the qubit-photon coupling strength. To achieve a good overlap between the final state and $|\text{NOON}\rangle$, it is essential to set $\lambda/\omega_a< 0.1$ and $\kappa/\omega_a < 0.1$. Under this condition, a sweeping speed in the order of $10^{-7}\omega_a^2$ can provide an adiabatic passage across the multiphoton resonance. In order to put our scheme into practice, the qubit decay rate and $Q$ factor of the resonator should be properly controlled.

From the above discussion, one immediately notices the advantage of realizing multiphoton resonances in the generalized Rabi model. For the original Rabi model and asymmetric Rabi model, only a single-photon coupling term exists. Then, at least $N-1$ intermediate states are involved to achieve a $N$-photon resonance. A simple perturbation theory analysis suggests that the gap at avoided crossing scales as $(\lambda/\omega_a)^N$. Since $\lambda/\omega_a\lesssim 0.1$, this high-order process is suppressed and cannot be effectively realized as in the gRM. For application, it becomes more challenging to realize adiabatic passage and generate NOON states with $N\geq 4$.

\section{Effects of multiphoton resonances on chiral transport in circuit QED lattices}
\label{sec:topo}

Recently, chiral quantum optics has become a burgeoning field that may find potential applications in quantum information and quantum communication~\cite{Lodahl-nature}. The propagation is said to be chiral if light travels in a single direction with a strong suppression of backscattering. It may also turn out that the absorption and emission of photons from an emitter depend on the propagation direction of light. Clearly, this type of chiral transport and light-matter interaction breaks time-reversal symmetry. One possible mechanism to achieve this is providing a tight confinement in the transverse direction of light, which develops into an emergent transverse spin. Then, the locking of spin and the propagation direction leads to the unidirectional flow of photons and chiral light-matter coupling~\cite{Nano-optics, Bliokh-Nori, SO-light, Aiello}. This task has become practical thanks to the tremendous progress in fabricating different nonreciprocal nanophotonic structures, such as photonic-crystal waveguides and optical nanofibers~\cite{Lee2012, Fortuno2013, Lin-science2013, Luxmoore-PRL, Junge2013, Luxmoore2013, Petersen2014, Neugebauer2014, Mitsch2014, ACS2014, Shomroni2014, Oconnor2014, Sollner2015, le-Feber2015, Coles2016}.

In addition to the above mechanism, chiral photon transport can be also achieved by inducing a synthetic gauge field for photons~\cite{Girvin-TRS, Girvin-JC, Carusotto2011, RMP-topo-optics, Hafezi2014, Hey-Li2018, nature-topo-optics, focus-NJP}. This approach can be better understood if one analogizes it to the quantum Hall effect in electronic systems. In the simplest case of an integer quantum Hall liquid, electrons propagate with a definite chirality on the edge of the system~\cite{Halperin-edge, Buttiker-edge}. This chiral flow of electrons originates from the breaking of time-reversal symmetry (TRS) and the nontrivial topology in the system due to the external magnetic field~\cite{TKNN, Avron-review}. A similar effect, known as the quantum anomalous Hall effect, may also occur in a lattice without any external magnetic field~\cite{Haldane-QAHE}. In this case, electrons hop in the lattice with a complex hopping amplitude, and the accumulated Aharonov-Bohm phase plays the role of an effective magnetic flux. With this knowledge, one may think of breaking TRS and achieving chiral transport in photonic lattices in a similar fashion. However, this task is not straightforward as photons are chargeless in an electromagnetic field. In order to put the scheme in practice, it was proposed to insert on-chip circulators made up of Josephson rings with enclosed magnetic fluxes in cQED lattices~\cite{Girvin-TRS}. By doing so, a complex phase factor in the photon hopping term is induced. At the end, both TRS breaking and chiral transfer of a Fock state with a single photon can be accomplished by tuning the phase factor properly. Based on the current development in cQED architecture, the proposal is argued to be accessible in real practice.

In the above discussion, light-matter interaction was completely left out of consideration. When the interaction is included, both the energy spectrum and the quantum dynamics of the system can be modified. Instead of trapped atoms and cavity photons~\cite{cavity-QED}, microwave resonators and superconducting qubits are the basic elements to simulate quantum optical processes in a cQED system~\cite{cQED-review}. Meanwhile, the system can be still described by the Rabi model and its generalization. By coupling the microwave resonators and their corresponding individual qubits as a light-matter interaction, photon transport in cQED lattices equipped with synthetic fluxes was examined in Ref.~\cite{Girvin-JC}. The previous study focused on the regime where the RWA applies, such that the system reduces to a Jaynes-Cummings lattice. Both quantum dynamics and the corresponding homodyne transmission amplitude of the system were found to depend sensitively on the coupling strength. This result is believed to have a profound impact on quantum measurement.

By taking counter-rotating processes into account, we demonstrated in previous sections a possible coupling between a two-level system and photons via multiphoton resonances. Since the ultrastrong-coupling regime can be reached effectively in a cQED system, this kind of resonant coupling is relevant and should not be neglected. Reviewing the system being investigated in Ref.~\cite{Girvin-JC}, it is unclear how photon transfer and multiphoton resonances will interplay with each other. Supposing there are $n>1$ photons in the system, one may ask if they can be still transferred chirally in the cQED lattice. This problem is not trivial as these photons can be absorbed to excite a qubit via an $n$-photon resonance. At the opposite extreme, one may also ask, can the photon-hopping process being suppressed? The answers to these questions will provide insight into chiral photon transport in the USC regime and complement previous work. More generally, the results may also suggest the importance of investigating the effect of light-matter interaction in topological photonic systems, in which chiral propagation of light may also be observed.

\begin{figure} [htb]
\includegraphics[width=2.5in]{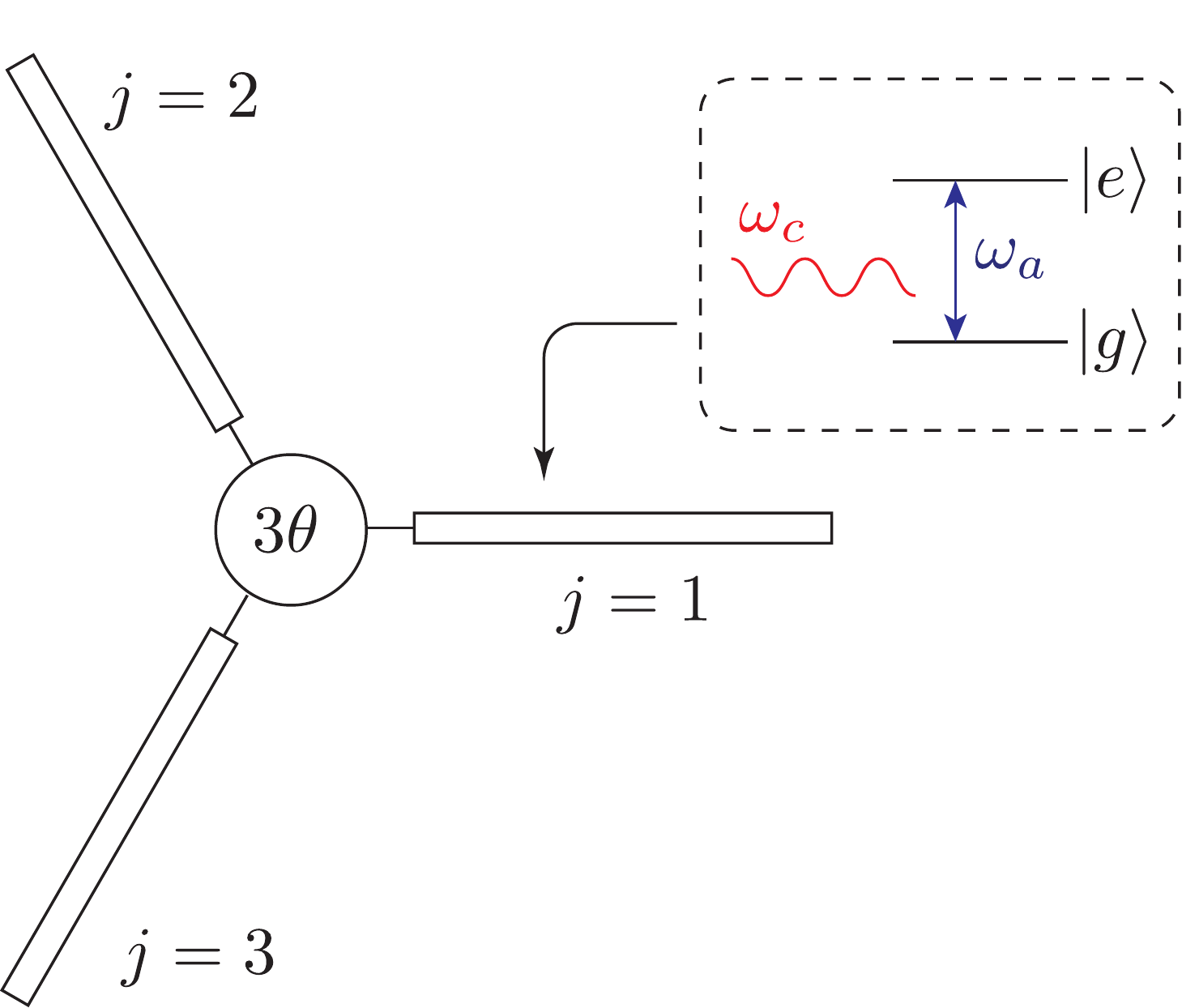}
\caption{(Color online) Schematic diagram of the system with a junction connected to three microwave resonators. Each resonator is coupled to a qubit as shown in the inset. Time-reversal symmetry is broken by introducing a synthetic magnetic flux in the system. An effective Aharonov-Bohm phase of $3\theta$ is gained when a photon hops around the system.}
\label{fig:setup}
\end{figure}

In the following discussion, we revisit the setup considered in Ref.~\cite{Girvin-JC}. To be specific, the system is a junction connected to three microwave resonators as shown in Fig.~\ref{fig:setup}. Each microwave resonator is coupled to a superconducting qubit. Here, we do not employ the RWA. We further assume that each qubit-resonator system is described by the generalized Rabi model in Eq.~\eqref{eq:Hamiltonian-gRM}. In addition, photons can hop between the neighboring resonators with a complex hopping amplitude $J'=Je^{i\theta}$. It was suggested that the TRS can be broken by choosing $3\theta\neq \pi\mathbb{Z}$~\cite{Girvin-TRS, Girvin-JC}. The breaking of TRS is necessary (but not sufficient) for realizing a chiral transfer of photons in the system. Based on the above discussion, the Hamiltonian describing the system is
\begin{align} \label{eq:system-Hamiltonian}
H=\sum_{j=1}^3 H^{\text{gRM}}_j
+J\sum_{j=1}^3 \left(a_{j+1}^\dagger a_j e^{-i\theta} + \text{H.c.}\right).
\end{align}
Here, the symbol $H^{\text{gRM}}_j$ denotes the generalized Rabi Hamiltonian for the $j$ th resonator. It takes the form in Eq.~\eqref{eq:Hamiltonian-gRM}. In addition, we tune the optical frequencies for all three resonators to the multiphoton-resonant frequency $\omega_c'$, as predicted from the perturbation theory approach in Sec.~\ref{sec:eqn-resonance}. The value of 
$\omega_c'$ depends on which multiphoton resonance one wants to study.

Before solving the Hamiltonian exactly, we can understand and predict some features of the short-time dynamics of the system. When the dynamics is completely dominated by photon hopping, it suffices to neglect the qubits and focus on the following Hamiltonian:
\begin{eqnarray} \label{eq:H_hop}
H_{\text{hop}}
=\sum_{j=1}^3 \omega_c a_j^\dagger a_j
+J\sum_{j=1}^3 \left(a_{j+1}^\dagger a_j e^{-i\theta} + \text{H.c.}\right).
\end{eqnarray}
This Hamiltonian can be easily diagonalized. Suppose the TRS is broken by choosing 
$\theta=\pi/6$. Then, a chiral photon transfer was predicted if the system has one photon in any one of the resonators initially~\cite{Girvin-TRS}. It requires a time $t_H=T_H/3=2\pi/(3\sqrt{3}J)$ for the photon to hop to the neighboring resonator. Here, $T_H$ is the period of photon hopping around the whole system. Different from the original work, our initial state has $n$ photons in one of the resonators. If the dynamics of the system is dominated by photon hopping, a chiral transfer of $n$ photons is expected.

On the other hand, one may consider the opposite limit by assuming the photon-hopping effect is negligible. Thus the photons are trapped in the same resonator. Suppose we choose an initial state: $\left|g,n\rangle\right.$ for the first resonator and $\left|g,0\rangle\right.$ for the other two resonators. In this case, a multiphoton resonance is expected in the first resonator due to the coupling between the bare states 
$|g,n\rangle$ and $|e,0\rangle$. When both $\lambda/\omega_a \ll 1$ and $\kappa/\omega_a \ll 1$ are satisfied, the multiphoton resonance can be approximated by the effective Hamiltonian in Eq.~\eqref{eq:reduced-H}. Then, the quantum state of the first resonator evolves approximately as
\begin{align} \label{eq:mp-oscillation}
\left|\psi(t)\rangle\right.
\approx e^{-iEt}
\left[
\cos{\left(\Omega_{\text{eff}}t\right)} \left| g,n\rangle\right.
+i\sin{\left(\Omega_{\text{eff}}t\right)} \left| e,0\rangle\right.
\right].
\end{align}
Here, $E=E_i+\Delta E_i=E_f+\Delta E_f$ for the two bare states with the energy corrected by Stark shift. The form of $\left|\psi(t)\rangle\right.$ suggests that a time interval 
$t_R=T_R/2=\pi/(2\Omega_{\text{eff}})$ is required for the qubit to be excited by the photons. We emphasize that both $E$ and $\Omega_{\text{eff}}$ are approximate results from perturbation theory. Thus Eq.~\eqref{eq:mp-oscillation} has neglected all possibly small but nonzero projections along other bare states. These projections also affects the quantum dynamics of the system, and are not negligible when the qubit-photon interaction becomes sufficiently strong. 

From the above discussion, one can define a scaleless parameter $\mu=t_R/t_H=3\sqrt{3}J/(4\Omega_{\text{eff}})$. Depending on $\mu$, the short-time dynamics of the system can transit from a chiral photon transfer to a suppression of photon transfer. In the following discussion, we use the four-photon resonance as a demonstration. The two extreme limits $\mu\gg 1$ and $\mu\ll 1$ are discussed separately.

\subsection{$\mu\gg 1$: chiral transfer of photons}
\label{sec:chiral-ph-flow}

We first consider the scenario when $t_R\gg t_H$. In this case, there is not enough time for the qubit to be excited before the photons are transferred to the next resonator. Hence, the short-time dynamics of the system is dominated by photon hopping. From the previous discussion, a chiral transfer of four photons and unexcited qubits are predicted. Nevertheless, the actual dynamics of the system will be modified by the qubit-photon interaction. 

\begin{figure}
\includegraphics[width=3.3in]{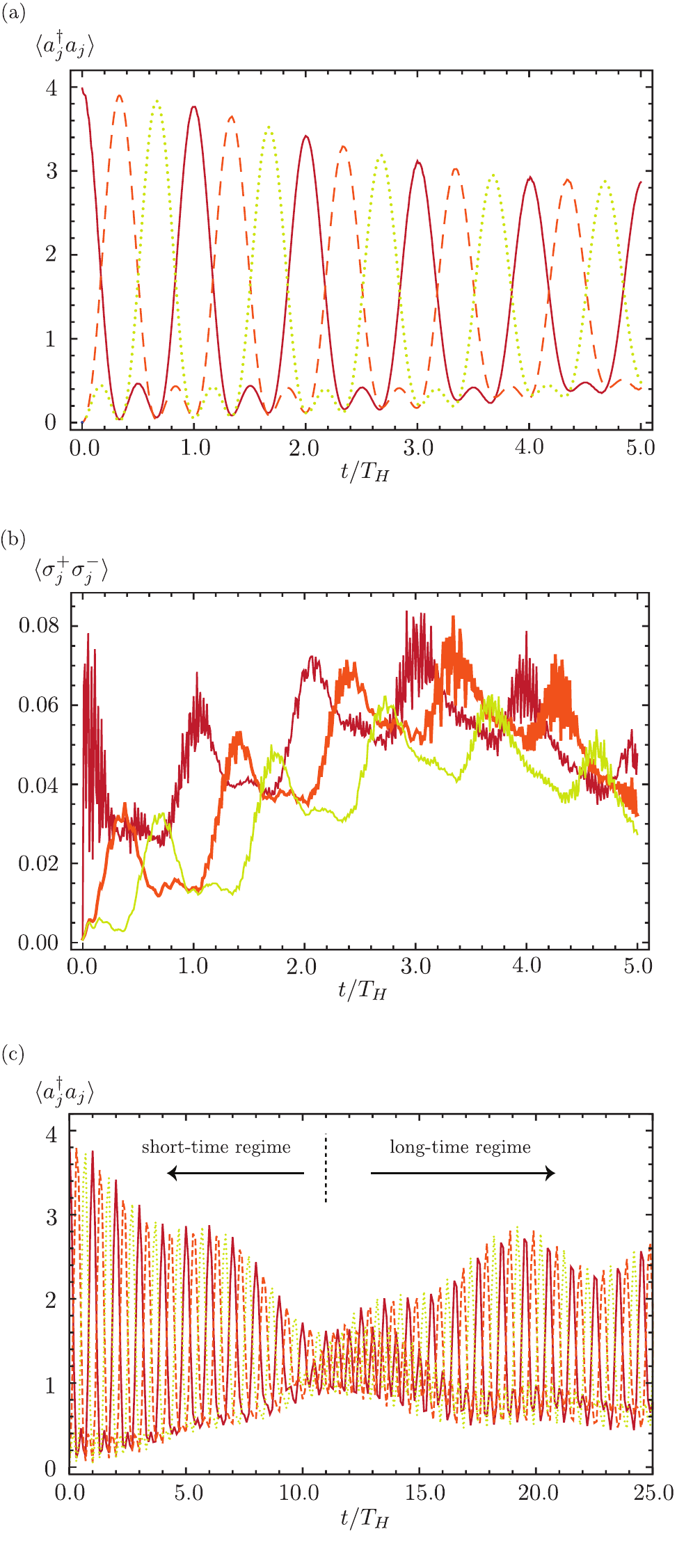} 
\caption{(Color online) Time evolution of the expectation values of (a) photon numbers $\langle a_j^\dagger a_j\rangle$ and (b) qubit excitations $\langle \sigma^+_j\sigma^-_j\rangle$ in the short-time regime. The transition out of the short-time dynamics is illustrated in (c). Here, the initial state of the system is $\left|\Psi(0)\rangle\right.=|g,4\rangle_1\otimes|g,0\rangle_2\otimes|g,0\rangle_3$. The solid red (darkest) line, dashed orange (lighter gray) line, and dotted yellow (lightest gray) line display the values for the first, second, and third resonators, respectively. Here, the parameters are $\omega_c/\omega_a\approx 0.258$, $\lambda/\omega_a=0.05$, $\kappa/\omega_a=0.01$, $\mu=10$, and $\theta=\pi/6$.}
\label{fig:4-ph-hop}
\end{figure}

We set $\lambda/\omega_a=0.05$ and $\kappa/\omega_a=0.01$ for the strengths of one-photon and two-photon terms (same parameters as in Fig.~\ref{fig:4-ph-av-crossing}). As a result, $\omega_c'\approx 0.258 \omega_a$ is predicted for the optical frequency to achieve a four-photon resonance. We set our initial state as $|\Psi(0)\rangle=|g,4\rangle_1 
\otimes |g,0\rangle_2\otimes|g,0\rangle_3$. By setting $\mu=10$ and $\theta=\pi/6$, we simulate the time evolution of the system numerically. Using the result, we evaluate both $\langle\Psi (t) | a_j^\dagger a_j |\Psi(t)\rangle$ and $\langle\Psi (t) | \sigma^+_j\sigma^-_j |\Psi(t)\rangle$. For convenience, these quantities are abbreviated as $\langle a_j^\dagger a_j\rangle$ and $\langle \sigma^+_j\sigma^-_j\rangle$ in the following discussion. The numerical results are shown in Fig.~\ref{fig:4-ph-hop}. 

For the short-time dynamics, the system shows a chiral transfer of photons between the resonators with a period of $T_H$. At the same time, $\langle a_j^\dagger a_j\rangle$ show decreasing peaks due to the modulation from four-photon Rabi oscillations. By increasing $\mu$, the peaks can attain values closer to $4$. Also, the chiral transfer of photons can persist for a longer period of time. Note that the eigenstates of the qubit-photon system at the multiphoton resonance are not perfectly given by $|\pm\rangle=(|e,0\rangle \pm |g,4\rangle)/\sqrt{2}$. There are small projections along other bare states, such as $|e,1\rangle$, $|e,2\rangle$, etc. The possibility of exciting the qubits to these bare states also contributes to the decreasing $\langle a_j^\dagger a_j \rangle$ and nonvanishing $\langle\sigma^+_j\sigma^-_j\rangle$. Their contributions are small and the corresponding oscillations should be much faster. 

To investigate the transition out of the short-time dynamics, we simulated the time-evolution of the system for $t \leq 50 T_H$. Within this period of time, $\langle\sigma^+_j\sigma^-_j\rangle$ remain small and do not go beyond $0.1$. It indicates that the qubits are nearly unexcited. For a better illustration, we only show $\langle a_j^\dagger a_j\rangle$ with $t\leq 25 T_H$ in Fig.~\ref{fig:4-ph-hop}(c). From the figure, we identify the occurrence of transition at $t\approx 11 T_H$. Although the amplitudes of $\langle a_j^\dagger a_j\rangle$ are decreasing, photons are transferred among the resonators chirally before the transition. After the transition, the chiral transfer becomes less obvious and less significant.

\subsection{$\mu\ll 1$: Multiphoton Rabi oscillation and suppression of photon transfer}
\label{sec:ph-suppression}

By changing $\mu$ to $0.1$ and keeping all parameters unchanged, we examine the dynamics in the opposite limit. The short-time dynamics of the system is dominated by multiphoton resonance. Once the rotating-wave approximation is made, there will be no multiphoton resonance (except the three-photon resonance when $\kappa\neq 0$). Therefore, the dynamics strongly depends on the counter-rotating processes in the generalized Rabi model.

\begin{figure} [htb]
\includegraphics[width=3.3in]{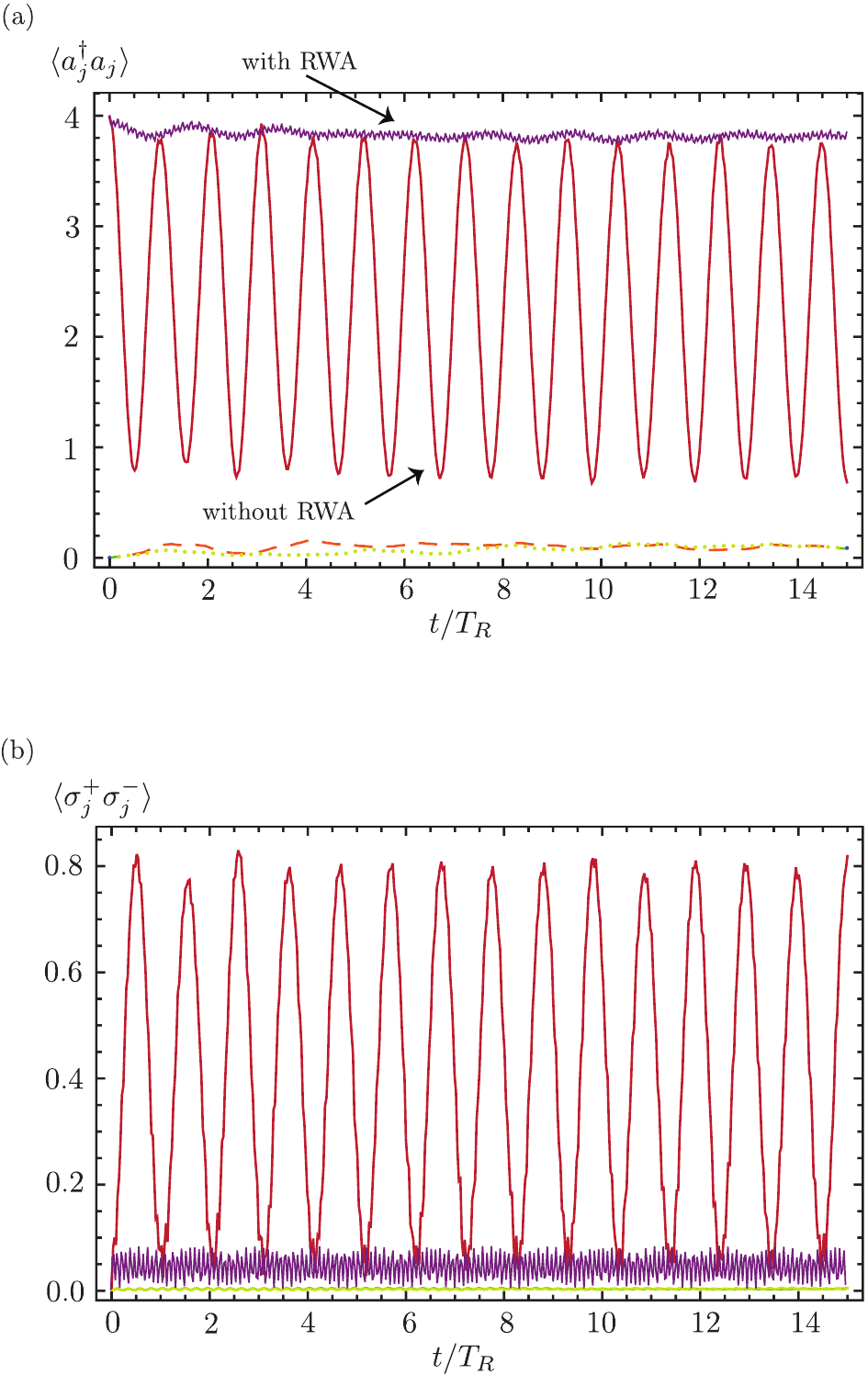} 
\caption{(Color online) Time evolution of the expectation values of (a) photon numbers $\langle a_j^\dagger a_j\rangle$ and (b) qubit excitations $\langle \sigma^+_j\sigma^-_j\rangle$ in the short-time regime. The solid red (darkest) line, dashed orange (gray) line, and dotted yellow (lightest gray) line display the values for the first, second, and third resonators, respectively. For comparison, the expectation values $\langle a_1^\dagger a_1\rangle$ and $\langle \sigma^+_1\sigma^{-}_1\rangle$ with the RWA are shown by the purple fuzzy lines. The same initial state of the system and parameters in Fig.~\ref{fig:4-ph-hop} are adopted, except $\mu$ is changed to $0.1$.}
\label{fig:4-ph-mp-res}
\end{figure}

Using the same initial state as before, we simulate the time evolution of the system for 
$0\leq t \leq T_H$. In the short-time regime, four-photon resonance in the first resonator is anticipated. Since $\mu=0.1$, we have $T_H=15T_R$. Hence, approximately $15$ four-photon Rabi oscillations between $|g,4\rangle$ and $|e,0\rangle$ should be observed. Our numerical results are shown in Fig.~\ref{fig:4-ph-mp-res} which support our prediction. Small projections along other bare states and the tiny probability of photon transfer out of the resonator make the multiphoton Rabi oscillation between $|g,4\rangle$ and $|e,0\rangle$ imperfect. We have numerically verified that this can be improved by tuning $\lambda\rightarrow 0$, $\kappa\rightarrow 0$, or reducing the photon hopping strength, i.e., $\mu\rightarrow 0$. This feature actually agrees with Ref.~\cite{gRM-solution-PRA}, which studied the gRM in the absence of photon hopping. When both $\lambda$ and $\kappa$ are nonzero, there would be two different Rabi frequencies in the system. Hence, their interplay makes it impossible to observe a perfect Rabi oscillation. Although neither the single-photon oscillation nor the two-photon oscillation is the most relevant in our present case, the simultaneous presence of two Rabi frequencies still makes the multiphoton Rabi oscillation imperfect. In addition, Fig.~\ref{fig:4-ph-mp-res} clearly confirms the absence of four-photon resonance under the RWA. Since the photon frequency is largely detuned from the transition frequency of the qubit, the probabilities of exciting the qubit by one-photon and two-photon co-rotating processes are small. This is reflected in the slight modulation of 
$\langle a_1^\dagger a_1\rangle$ and the small-amplitude rapid oscillation in 
$\langle \sigma^+_1\sigma^-_1\rangle$ when the RWA is applied. 

\begin{figure} [htb]
\includegraphics[width=3.3in]{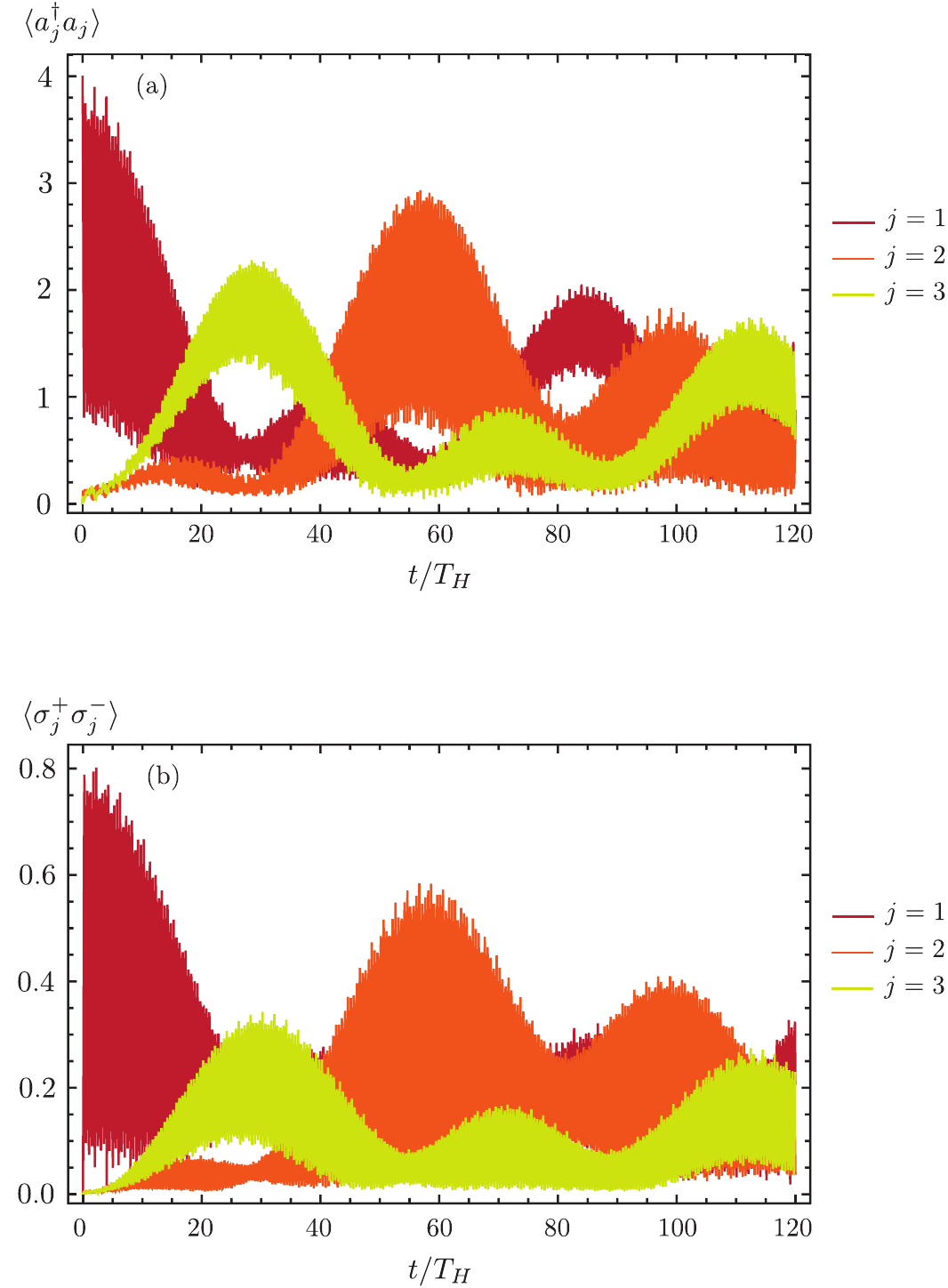} 
\caption{(Color online) Time evolution of the expectation values of (a) photon numbers $\langle a_j^\dagger a_j\rangle$ and (b) qubit excitations $\langle \sigma^+_j\sigma^-_j\rangle$ in a longer time regime. The same initial state of the system and parameters in Fig.~\ref{fig:4-ph-mp-res} are adopted.}
\label{fig:4-ph-mp-res-long}
\end{figure}

Another important feature in the short-time dynamics is the small magnitudes of 
$\langle a_2^\dagger a_2\rangle$ and $\langle a_3^\dagger a_3\rangle$. Naively, one may expect photon transfer across the resonator junction would occur within a time scale $t\gtrsim T_H$. However, our numerical result for the dynamics in a longer time regime in Fig.~\ref{fig:4-ph-mp-res-long}  disproves the idea. The figure shows that $\langle a_2^\dagger a_2\rangle$ and $\langle a_3^\dagger a_3\rangle$ remain small even when $t\approx 10 T_H$. Instead of the original time scale $t_H=T_H/3$, a much longer time $t\approx 30 T_H$ is required for photon transfer between different resonators. In other words, the process is strongly suppressed by the four-photon resonance in the first resonator. At the same time, there is no preferred chirality in the photon transfer.

\subsection{Possible implication for energy transfer}

Suppose the first and the third resonators in Fig.~\ref{fig:setup} are coupled to heat baths with temperatures $\Theta_h$ and $\Theta_l$, respectively. This coupling introduces dissipation from photon leakage and qubit decay, with the respective decay rates $\gamma_p$ and $\gamma_a$. The nonzero temperature $\Theta$ of the heat bath modifies the effective decay rates by the Bose-Einstein distribution $n_B(\Theta)$. Here, the dissipation is assumed to be weak such that 
$\left[n_B(\Theta)\gamma_p\right]^{-1}, \left[n_B(\Theta)\gamma_a\right]^{-1} \gg \text{min }(T_H, T_R)$. Consider the case when $\Theta_h>\Theta_l$, for which heat and energy flow in the system. This flow of energy relies on the photon-hopping process between different resonators. Based on the previous results, we discuss qualitatively possible implication for energy transfer in the system.

The nonequilibrium distribution of photon number in each resonator can be very complicated. However, the chiral transfer of photons is affected only when the qubit frequency and photon frequency match the condition of multiphoton resonance. Supposing the condition is satisfied, a chiral transfer of photons is still expected when $\mu\gg 1$. Thus we expect heat flows in the system with a preferred chirality, the same as the photon hopping discussed in Sec.~\ref{sec:chiral-ph-flow}. On the other hand, a similar argument and the result in Sec.~\ref{sec:ph-suppression} suggest the suppression of energy flow in the $\mu\ll 1$ regime. A quantitative analysis can be performed by employing the generalized master equation developed in Ref.~\cite{Nori-dissipation}. This will be addressed elsewhere. 

Furthermore, it is tempting to study the effect of photon-atom interaction on energy transport in topological photonic systems. For example, a chiral flow of thermal current was reported in a Hofstadter square lattice with bosons~\cite{topo-heat-square}. This flow was shown to be robust against disorder, and stemmed from a topological protection. It is unclear if this chiral flow persists or not when atoms and photon-atom coupling are introduced to the lattice. As a first guess, our results may simply suggest that the chiral flow persists whenever the period of photon hopping is much shorter than the period of Rabi oscillation from the resonant coupling.

\section{Conclusions}  \label{sec:conclusion}

To conclude, we have investigated different multiphoton resonances in the generalized Rabi model (gRM) at $\omega_c\approx \omega_a/N$. We have focused on three-photon to six-photon resonances because they involve only two intermediate states. This feature allowed us to apply third-order perturbation theory and derive an effective Hamiltonian to describe the resonance. In particular, we have obtained an analytic approximation for the resonant frequency $\omega_c'/\omega_a$ and energy splitting $\Delta/\omega_a$ of the avoided crossing. The validity of the approximation has been checked explicitly for three-photon and four-photon resonances. By comparing to the results from numerical diagonalization, we found that both approximate $\omega_c'/\omega_a$ and $\Delta/\omega_a$ have errors less than $10\%$, when $\lambda/\omega_a<0.05$ and $\kappa/\omega_a<0.05$. 

Our work has highlighted the versatility and effectiveness of realizing multiphoton resonances in the gRM. They originate from the additional two-photon term in the model. We have provided a concrete example, showing that the lack of parity symmetry enabled us to achieve four-photon and six-photon resonances. Furthermore, the lowest order coupling between 
$|g,N\rangle$ and $|e,0\rangle$ via counterrotating processes in the asymmetric Rabi model is an $N$ th-order process. Therefore, $\Delta/\omega_a\sim (\lambda/\omega_a)^N$ for the corresponding $N$-photon resonance. In contrast, our study showed that a third order dependence holds for all three- to six-photon resonances in the gRM. For a typical circuit QED setup operating in the ultrastrong regime, both $\lambda/\omega_a$ and $\kappa/\omega_a$ can reach the order of $0.1$. It is believed that all these resonances are significant and achievable in current experiments. We have further exploited this advantage and outlined a possible application of our results in NOON state generation. Our proposal relies on an adiabatic sweeping of the photon frequency across the multiphoton resonance. A large splitting of the energy levels at the resonance is preferred. The possibility of achieving high-number photon resonances via a lower-order coupling scheme makes it feasible to realize them in a real circuit QED setup. It is beyond the scope of the current paper to include a thorough numerical study of the proposal and include dissipative effects.

In addition, we have revisited the setup of a junction of three resonators coupled to their individual qubits~\cite{Girvin-JC}. By not employing the rotating-wave approximation, it enabled us to realize multiphoton resonances in the system and examine its interplay with photon transport. Specifically, we quantified the competition between multiphoton resonances and the photon hopping effect by defining the ratio $\mu=3\sqrt{3}J/(4\Omega_{\rm eff})$. We chose $\omega_c\approx\omega_a/4$, and studied the short-time dynamics of the system at $\mu=10$ and $\mu=0.1$ numerically. The system underwent a transition from chiral transfer of four photons in the former case to a suppression of photon transfer in the latter case. This behavior is similar to the scenario when the system contains one photon only. In general, we expect the same transition can occur whenever $\omega_c\approx\omega_a/N$.

Finally, we believe our work has further motivated future studies of the generalized Rabi model. On the application side, our work may shed light on quantum measurement and manipulation of energy transport in cQED systems in the ultrastrong coupling regime. A concrete example comes from quantum nondemolition measurements (QNMs)~\cite{Girvin-QNM}, which rely on a dispersive interaction. In this situation, multiphoton resonances should be avoided. Given that both photon hopping and qubit-photon interaction strengths are tunable in realistic circuit QED setups, our results on chiral transport are relevant for real practice. For future work, it will be tempting to examine how chiral light transport in different topological photonic systems is affected by light-matter interaction.

% ------------------------------ Acknowledgements -----------------------------------------

\begin{acknowledgments}

The author gratefully acknowledges D. E. Feldman for his continuous support and comments on the manuscript. The author would also thank Zekun Zhuang, L. Cong, and Z.-J. Ying for many useful suggestions. This work was supported by the National Science Foundation under Grant No. DMR1902356 and the Galkin Foundation Fellowship under the Department of Physics at Brown University.

\end{acknowledgments}

\appendix

\section{An alternative derivation of effective Hamiltonians for multiphoton resonances}
\label{app:effective-H}

In this Appendix, we derive the effective Hamiltonian for multiphoton resonances in the generalized Rabi model by employing the James' effective Hamiltonian approach~\cite{effective-James}. By doing so, we verify the resonant frequency and effective coupling strengths in the main text. 

We start by rewriting the gRM Hamiltonian in the interaction picture with 
$H_0=(\omega_a/2)\sigma_z+\omega_c a^\dagger a$. From the Heisenberg equations of motion, i.e., $d\hat{O}/dt=i [H_0, \hat{O}]$, we have
\begin{align}
& a(t) = a e^{-i\omega_c t},
\\
& a^2(t) = a^2 e^{-2i\omega_c t},
\\
& \sigma_+ (t) = \sigma_+ e^{i\omega_a t}.
\end{align}
Using the above results, it is straightforward to deduce that
\begin{align}
\nonumber
H_I(t)
=&~\lambda\left[a e^{i \left(\omega_a-\omega_c\right)t} 
+a^\dagger e^{i\left(\omega_a+\omega_c\right)t}\right]\sigma_+ +\text{H.c.}
\\
&+\kappa\left[a^2 e^{i \left(\omega_a-2\omega_c\right)t} 
+\left. a^\dagger\right.^2 e^{i\left(\omega_a+2\omega_c\right)t}\right]\sigma_+ +\text{H.c.}
\end{align}
Here, $H_I(t)$ denotes the Hamiltonian for the generalized Rabi model in the interaction picture. For the $n$-photon resonance, we set $\omega_a=n\omega_c$. Then, we have
\begin{align}
\nonumber
H_I(t)
=&~\lambda\left[a e^{i \left(n-1\right)\omega_c t} 
+a^\dagger e^{i\left(n+1\right)\omega_c t}\right]\sigma_+ +\text{H.c.}
\\
&+\kappa\left[a^2 e^{i \left(n-2\right)\omega_c t} 
+\left. a^\dagger\right.^2 e^{i\left(n+2\right)\omega_c t}\right]\sigma_+ +\text{H.c.}
\end{align}

Based on the James' effective Hamiltonian approach, the effective Hamiltonian for multiphoton resonance can be obtained as~\cite{effective-James}
\begin{eqnarray}
H_{\text{eff}}(t)
=H^{(2)}_{\text{eff}}(t)+H^{(3)}_{\text{eff}}(t)+\cdots
\end{eqnarray}
Since we only focus on multiphoton resonances involving two intermediate states, a third-order perturbation theory is sufficient. Explicitly, the second-order and third-order correction terms are
\begin{align}
H_{\text{eff}}^{(2)}
=&~\frac{1}{i} H_I(t) \int^t H_I(t') dt',
\\
H_{\text{eff}}^{(3)}
=&~-H_I(t) \int^t  H_I(t_1)  \int^{t_1} H_I(t_2) dt_2 dt_1.
\end{align}

For convenience, we denote the photon-number operator $\hat{N}=a^\dagger a$. By only keeping terms which do not have oscillating phase factors, we obtain
\begin{align} \label{eq:second-eff}
\nonumber
H_{\text{eff}}^{(2)}
=&~\frac{\lambda^2}{(n^2-1)\omega_c}
\left[2n \hat{N} +\left(n+1\right)\right]\sigma_{+} \sigma_{-}
\\ \nonumber
&-\frac{\lambda^2}{(n^2-1)\omega_c}
\left[2n \hat{N} +\left(n-1\right)\right]\sigma_{-}\sigma_{+}
\\ \nonumber
&+\frac{2\kappa^2}{(n^2-4)\omega_c}
\left[(n+2)+(n+4) \hat{N} +n \hat{N}^2\right] \sigma_+ \sigma_{-}
\\
&-\frac{2\kappa^2}{(n^2-4)\omega_c}
\left[(n-2)+(n-4) \hat{N} +n \hat{N}^2\right] \sigma_{-} \sigma_+ 
\end{align}

For $H_{\text{eff}}^{(3)}$, the nonoscillating terms depend on the value of $n$. For example, we choose $n=3$ and obtain the third-order effective Hamiltonian for the three-photon resonance:
\begin{align}
\nonumber
H_{\text{eff}}^{(3)}
=&-\left(\frac{\lambda^3 \left.a^\dagger\right.^3}{4\omega_c^2}
+\frac{\kappa^2\lambda \left.a^\dagger\right.^2 a \left.a^\dagger\right.^2}{\omega_c^2}
+\frac{\kappa^2\lambda \left.a^\dagger\right.^4 a}{4\omega_c^2}
\right)\sigma_{-}
\\
&-\left(\frac{\lambda^3 a^3}{4\omega_c^2}
+\frac{\kappa^2\lambda a^2 a^\dagger a^2}{\omega_c^2}
+\frac{\kappa^2\lambda a^4 a^\dagger}{4\omega_c^2}
\right)\sigma_{+}.
\end{align}
By choosing different values of $n$, one may also obtain $H_{\text{eff}}^{(3)}$ for the four-, five-, and six-photon resonances. All results here are consistent with the calculation by Eq.~\eqref{eq:perturb-eff} in the main text.

\subsection{Resonant frequency}

Consider the reduced Hilbert space formed by bare states $\left|e,n_0\rangle\right.$ and 
$\left|g,n_0+n\rangle\right.$. Then, Eq.~\eqref{eq:second-eff} gives a $2\times 2$ diagonal Hamiltonian. An effective Hamiltonian can be obtained by rotating the result back to the laboratory frame. Then, the resonant frequency for the $n$-photon resonance between the bare states $\left|e,n_0\rangle\right.$ and $\left|g,n_0+n\rangle\right.$ is derived by equating the two diagonal matrix elements:
\begin{align} \label{eq:general-freq}
\nonumber
\left(\frac{\omega_c'}{\omega_a}\right)^{\text{$n$-ph}}
=&~\frac{1}{n} 
+\frac{2n(2n_0+n+1)}{n^2-1}\left(\frac{\lambda}{\omega_a}\right)^2
\\
&+\frac{2n\left[n_0^2+(n_0+n)^2+2n_0+n-2\right]}{n^2-4}
\left(\frac{\kappa}{\omega_a}\right)^2.
\end{align}
Physically, the quadratic correction terms come from the Stark shift in the energy levels. All equations for the resonant frequencies (without the RWA) in the main text can be reproduced from Eq.~\eqref{eq:general-freq}. Similarly, one can deduce the resonant frequency for the three-photon resonance between $\left|e,n_0\rangle\right.$ and $\left|g,n_0+3\rangle\right.$ under the RWA:
\begin{align}
\nonumber
\left(\frac{\omega_c'}{\omega_a}\right)^{\text{3-ph}}_{\text{RWA}}
=~&\frac{1}{3}
+\left(n_0+2\right)\left(\frac{\lambda}{\omega_a}\right)^2
\\
&+2\left(n_0+2\right)^2 \left(\frac{\kappa}{\omega_a}\right)^2.
\end{align}
By setting $n_0=0$, Eq.~\eqref{eq:freq-3-RWA} in the main text is reproduced.


\begin{thebibliography}{99}
\bibitem{footnote-Rabi-Stark}
Another related nonlinear model, the Rabi-Stark model was also introduced and studied recently~\cite{Rabi-Stark-intro, Rabi-Stark-p1, Rabi-Stark-p2}. Multiphoton resonances were discussed in the model and shown to be selective due to the additional Stark coupling term~\cite{Cong-selective}.
\end{thebibliography}
\end{document}